%% file: paper.tex
\documentclass[sigconf]{acmart}
\usepackage{subfigure}

\AtBeginDocument{%
  \providecommand\BibTeX{{%
    \normalfont B\kern-0.5em{\scshape i\kern-0.25em b}\kern-0.8em\TeX}}}

\copyrightyear{2022} 
\acmYear{2022} 
\setcopyright{acmcopyright}
\acmConference[SIGIR '22]{Proceedings of the 45th International ACM SIGIR Conference on Research and Development in Information Retrieval}{July 11--15, 2022}{Madrid, Spain}
\acmBooktitle{Proceedings of the 45th International ACM SIGIR Conference on Research and Development in Information Retrieval (SIGIR '22), July 11--15, 2022, Madrid, Spain}
\acmPrice{15.00}
\acmDOI{10.1145/3477495.3531785}
\acmISBN{978-1-4503-8732-3/22/07}

\usepackage{braket}
\usepackage{booktabs}
\usepackage{multirow}
\usepackage{makecell}

\begin{document}
\fancyhead{}
\title{Denoising Time Cycle Modeling for Recommendation}

\author{Sicong Xie}
\affiliation{
  \institution{Ant Group}
  \city{Hangzhou}
  \state{Zhejiang}
  \country{China}
  }
\email{sicong.xsc@antgroup.com}

\author{Qunwei Li}
\affiliation{
  \institution{Ant Group}
  \city{Hangzhou}
  \state{Zhejiang}
  \country{China}
  }
\email{qunwei.qw@antgroup.com}

\author{Weidi Xu}
\affiliation{
  \institution{Ant Group}
  \city{Shanghai}
  \country{China}
  }
\email{weidi.xwd@antgroup.com}

\author{Kaiming Shen}
\affiliation{
  \institution{Ant Group}
  \city{Beijing}
  \country{China}
  }
\email{kaiming.skm@antgroup.com}

\author{Shaohu Chen}
\affiliation{
  \institution{Ant Group}
  \city{Beijing}
  \country{China}
  }
\email{shaohu.csh@antgroup.com}

\author{Wenliang Zhong}
\affiliation{
  \institution{Ant Group}
  \city{Hangzhou}
  \state{Zhejiang}
  \country{China}
  }
\email{yice.zwl@antgroup.com}

\renewcommand{\shortauthors}{Xie, et al.}

\begin{abstract}
Recently, modeling temporal patterns of user-item interactions have attracted much attention in recommender systems. We argue that existing methods ignore the variety of temporal patterns of user behaviors. We define the subset of user behaviors that are irrelevant to the target item as noises, which limits the performance of target-related time cycle modeling and affect the recommendation performance. In this paper, we propose Denoising Time Cycle Modeling (DiCycle), a novel approach to denoise user behaviors and select the subset of user behaviors that are highly related to the target item. DiCycle is able to explicitly model diverse time cycle patterns for recommendation. Extensive experiments are conducted on both public benchmarks and a real-world dataset, demonstrating the superior performance of DiCycle over the state-of-the-art recommendation methods.
\end{abstract}

\ccsdesc[500]{Information systems~Recommender systems}
\ccsdesc[500]{Information systems~Personalization}
\ccsdesc[500]{Information systems~Retrieval models and ranking}

\keywords{Time cycle, denoise, user behaviors, recommendation}


\maketitle
\section{Introduction}
\input{intro.tex}

\section{Related Work}
\input{related_work.tex}

\section{Denoising Time Cycle Modeling}
\input{model.tex}

\section{EXPERIMENTS}
\input{experiment}

\section{CONCLUSION}
In this paper, we proposed DiCycle to capture dynamic time cycle patterns. Combined with the gated filter unit, DiCycle can filter out the noises in user behaviors, which is defined as the subset of user behaviors that are irrelevant to the target item. DiCycle is able to model both absolute time pattern and relative time pattern in user behaviors which have high relevance to the target item. Extensive experiments on public benchmarks and a real-world dataset verified the effectiveness of DiCycle.
\newpage
\bibliographystyle{ACM-Reference-Format}
\bibliography{reference}
\end{document}

%% file: intro.tex
Personalized recommender systems can capture dynamic user interests and potential demands through analyzing a large amount of features, such as user profile, user behaviors, item category, etc~\cite{guo2017deepfm,lian2018xdeepfm,xiao2017attentional}. 
Among them, temporal information in user behaviors plays a vital role, which is extracted from user-item interaction records and can present different time-related patterns of user behaviors.

\begin{figure}[t]
\centering
\subfigure[]{
\includegraphics[width=40mm,height=28mm]{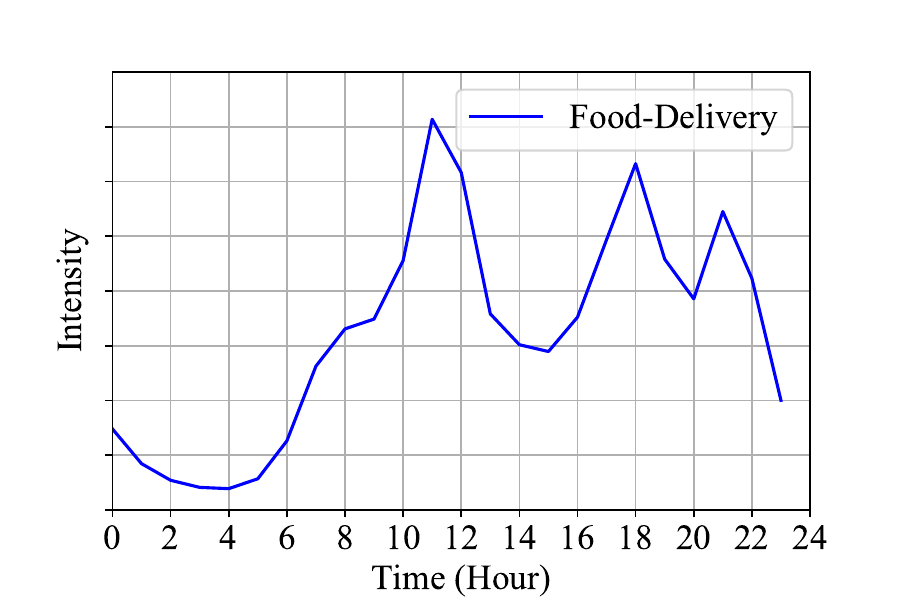}
}
\subfigure[]{
\includegraphics[width=40mm,height=28mm]{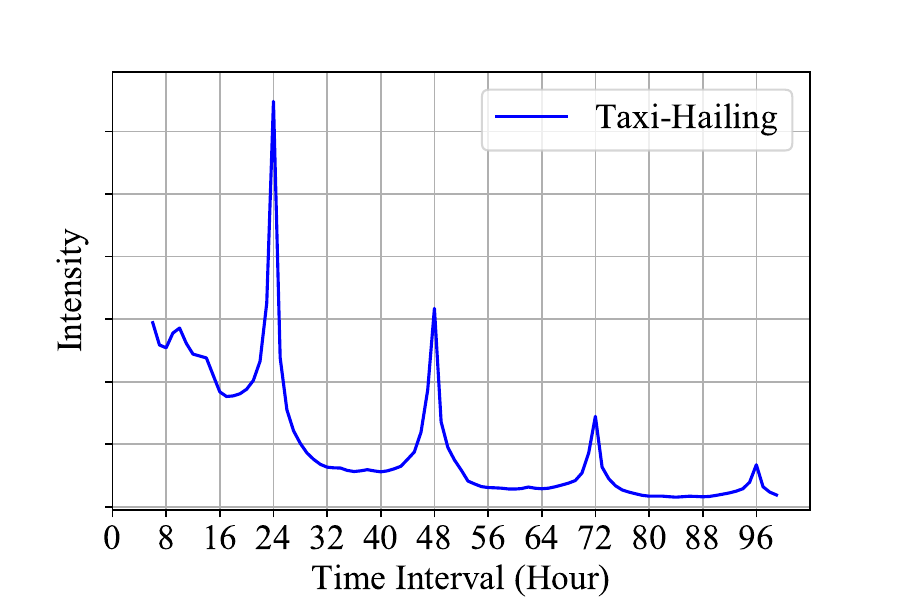}
}
\caption{
Analysis of two typical temporal patterns of user behaviors. \label{fig_taxi_delivery}}
\end{figure}

We analyze user's click actions on two items in our real-world dataset to illustrate two typical temporal patterns. Figure~\ref{fig_taxi_delivery}(a) shows how the number of clicks on an item of Food-Delivery service changes every hour in a day. 
The click intensity reaches its peaks at 11, 18, and 21 o'clock, corresponding to probable lunch, dinner, and midnight meal time. 
We call such a time pattern as Absolute Time Cycle Pattern (ATC), which characterizes the impact of semantic time (e.g., hour, weekday, day of month, etc) on user-item interactions and reveals the specific time points that a user may be interested in an item. 
Besides, the time span $|t_{2}-t_{1}|$ of timestamp $t_{1}$ and $t_{2}$ of user behaviors reveals critical temporal information as well.
Figure~\ref{fig_taxi_delivery}(b) shows that user's click intensity on an item of Taxi-Hailing service varies periodically after user's last click, which we refer to as the Relative Time Cycle Pattern (RTC). 
Such an observation hints that people tend to hail a taxi again 24 hours later after the last click and the intensity declines as time goes on, but still reaches local peaks every 24 hours. 
RTC directly illustrates how user's past behavior influences their current intentions after a certain time interval.

Existing works have studied the dynamic temporal patterns of user behaviors and proposed elaborate model structures \cite{li2020time,cho2021learning,ye2020time}.
However, these studies model time information of user behaviors of all the items without carefully differentiation and some items might be irrelevant to the target item.
For example, when the target is an item of Taxi-Hailing service, the user's past behaviors on taxi-related items would benefit the modeling of time cycle patterns and improve recommendation performance. On the contrary, past behaviors on completely irrelevant items (e.g., items of E-Commerce service) may bring adverse impact on time cycle modeling, which are defined as the noises of user behaviors given a specific target item in our paper. Therefore, denoising user behaviors is important for capturing target-related time cycle patterns and improving the recommendation performance.

In this paper, we propose Denoising Time Cycle Recommendation (DiCycle), a framework that disentangles noises in user behaviors and jointly learns the two time cycle patterns, i.e., ATC and RTC.
Representing the time has always been a challenge in machine learning.
As for ATC, the semantic time such as hour, day, and week, is a discrete variable, and representation of discretized time values can lose some information. In DiCycle, a convolutional module is applied to aggregate the adjacent time representation to mitigate the information loss caused by time discreteness. 
As for RTC, we utilize a continuous translation-invariant kernel to convert timestamp into embedding representations motivated by Bochner’s Theorem~\cite{xu2019self,xu2020inductive}. Based on this kernel, the inner product of the embeddings of timestamp $t_{1}$ and $t_{2}$ can characterize the relevance of user's behaviors on $t_{1}$ and $t_{2}$, depending on the time interval $|t_{2}-t_{1}|$.
Moreover, a gated filter unit softly selects a subset from user behaviors based on its relevance to the target item, filtering out noises for time cycle modeling. Combined with two temporal representations of ATC and RTC, a time cycle attention is applied to generate the final representation, indicating whether the user will click the target item from the perspective of time cycle pattern.

We verify DiCycle on three public benchmarks and a real-world dataset from our company.
The results show the proposed method can effectively capture diverse time cycle patterns of user behaviors and prominently improve the recommendation performance.

%% file: related_work.tex
Recently, there have been great advances in modeling dynamic patterns of user behaviors in deep learning, including recursive neural network (RNN)~\cite{tan2016improved}, convolutional neural network~\cite{tang2018personalized,xu2019recurrent, yan2019cosrec} and attention mechanism based structures~\cite{kang2018self,zhou2018atrank,zhou2018deep,zhou2019deep}.
These methods consider the sequential nature of user behaviors, processing user behavior sequences with natural language processing techniques.
Session-based approaches separate user behaviors according to their occurring time and model user short-term preferences, indirectly incorporating temporal information~\cite{chen2018sequential,feng2019deep}.
More recent researches directly focus on modeling temporal information, demonstrating its importance on many tasks, such as sequence prediction, and click-through rate (CTR) prediction~\cite{zhu2017next,xu2019self,xu2020inductive,ye2020time,cho2021learning}. 
TimeLSTM equips LSTM with time gates to model time intervals, which can better capture both user's short-term and long-term interests~\cite{zhu2017next}.
Several time-aware frameworks are specially designed for recommender systems, which learns heterogeneous temporal patterns of user preference~\cite{ye2020time,cho2021learning,costa2019collective,li2020time}.
However, how to extract denoised temporal information from user behavior for end-to-end applications as described above remains a challenge. The proposed DiCycle attempts to solve this problem, which is introduced in detail in the next section.

%% file: model.tex
\begin{figure*}
    \centering
    \includegraphics[width=17.8cm]{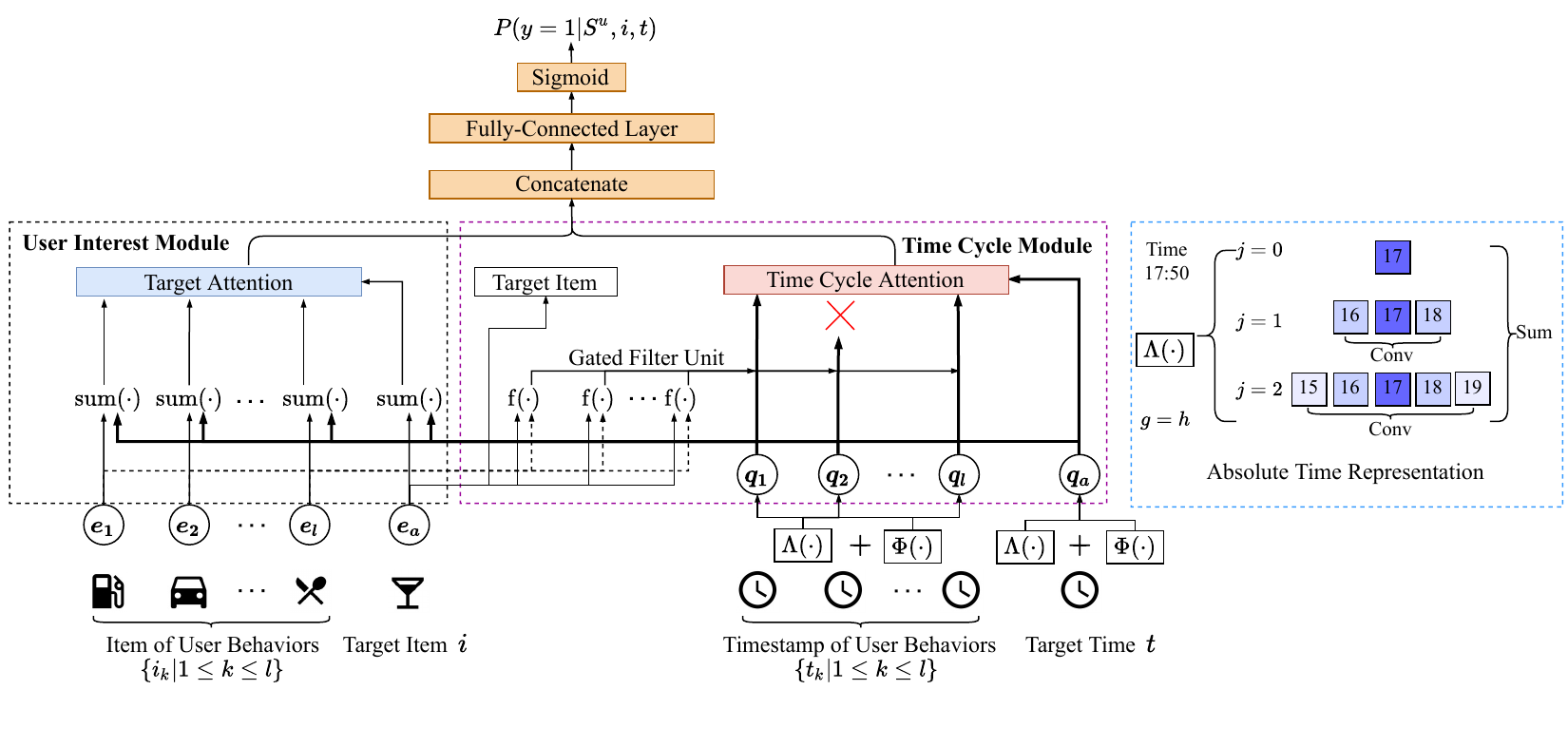}
    \caption{The overall architecture of the proposed {model}.}
    \label{fig-model}
\end{figure*}

In this section, we first define the problem formally, and then introduce the approach of time encoding for both ATC and RTC. 
Together with a gated filter unit to denoise user behaviors, we finally propose DiCycle for both ATC and RTC, the overall structure of which is shown in Figure~\ref{fig-model}.

\subsection{Problem Formulation}
Given a user $u\in\mathcal{U}$, a target item $i$ $\in\mathcal{I}$ and target timestamp $t$, our CTR task is to predict the probability that user $u$ will click item $i$ at timestamp $t$ as $P(y=1|u,i,t)$, where $y=1$ refers to click and $y=0$ refers to non-click. 
The target behavior to be predicted is defined as $s_{a}^{u}=(i,t)$.
Behaviors of user $u$ before time $t$ are organized as $S^{u}=\{s^{u}_{k}|1\leq {k}\leq{l^{u}}\}=\{(i^{u}_{k},t^{u}_{k})|1\leq {k}\leq{l^{u}}\}$, where $(i_{k}^{u},t_{k}^{u})$ means that user $u$ clicked on item $i_{k}^{u}$ at time $t_{k}^{u}$, and $t_{1}^{u}\leq t_{2}^{u}\leq \cdots  \leq t_{l_{u}}^{u} \leq t$ holds. 
Thus the CTR probability in our task turns to $P(y=1|S^{u},i,t)$. 
In the following, we omit $u$ in the notation for convenience.

\subsection{Time Encoding Unit}
In this section, we introduce the time encoding unit, including the representation learning of absolute time and of relative time 
\subsubsection{Learning Absolute Time Representation for ATC} 
Normally, for timestamp $t$, an embedding layer is applied to encode it into semantic time embeddings with different granularities, containing hour $h$, weekday $w$, month $m$ and etc, which can be formulated as $\Lambda_{g}(t)\in{\mathbb{R}^{d}}$ for granularity $(g\in G={\{h, w, m, \cdots\}})$. However, time continuity is broken in such a modeling. For example, the time point of 17:50 and 18:10 is quite close, but the hour time is 17 and 18, respectively, whose semantic embedding could be entirely different.
To combat such a loss of continuity of time, we propose an absolute time convolution module, which symmetrically incorporates surrounding time slots. Take time 17:50 and the granularity of hour for an example. Instead of just using the information of hour 17, its embedding by the granularity of hour is combined with the embedding of its surrounding hours, i.e., hour 15, 16, 17, 18, 19 and etc. Formally, the surrounding time embeddings of time $t$ by granularity $g$ is denoted as $\Lambda_{g,j}^{S}(t) = [\Lambda_{g}(t^{-j}), \cdots, \Lambda_{g}t^{0}),\cdots, \Lambda_{g}(t^{+j})]^{\mathrm T}\in{\mathbb{R}^{(2j+1)\times d}}$, where $j$ denotes the surrounding range of the target time and $\Lambda_{g}(t^{0})=\Lambda_{g}(t)$ holds. 

With the convolution kernel $K \in{\mathbb{R}^{n\times{d}}}$ and the activation function ReLU, where $n$ is the kernel size, we conduct one-dimensional convolution on $\Lambda_{g,j}^{S}(t)$ to get $\Lambda_{g,j}^{C}(t)\in{\mathbb{R}^{(2j+1)\times d}}$. Subsequently, max-pooling is conducted on the first dimension of $\Lambda_{g,j}^{C}(t)$ to get $\Lambda_{g,j}^{P}(t)\in{\mathbb{R}^{d}}$. CNN has the capability of capturing local features, which enables us to capture the similarities in surrounding time and maintain some time continuity.

As is shown in Figure~\ref{fig-model}, we set the surrounding range $j$ from $0$ to $J$ ($J=2$ in default), based on which the impact of the surrounding time gradually diminishes as it goes larger. The final representation of absolute time $\Lambda(t)$ is composed of all the granularities as



\begin{equation}
\begin{aligned}
\Lambda(t) = \sum_{g\in{G}}\sum_{j=1}^{J} {\rm Pooling} \left({\rm ReLU} \left({\rm Conv} \left(\Lambda_{g,j}^{S}(t), K\right)\right)\right) \,.
\end{aligned}
\end{equation}

\subsubsection{Learning Relative Time Representation for RTC}
Relative time is defined as the time interval between the target timestamp $t$ and the timestamp $t_{k}$ in user's past behaviors, which reflects how much influence her past behaviors are on current intention. We aim to find a mapping function $\Phi$ that transforms time interval from time domain $\mathcal{T}$ to $d-$dimensional vector space, preserving the evolving nature of time information. In this way, the information of time interval $|t_{2}-t_{1}|$ of timestamps $t_{1}$ and $t_{2}$ can be extracted by the inner product $<\Phi(t_{1}),\Phi(t_{2})>$. Therefore, the objective turns to learning a translation-invariant temporal kernel $\mathcal{K} (t_{1},t_{2})=\psi(t_{2}-t_{1})=<{\Phi}(t_{1}),{\Phi}(t_{2})>$, where $\psi:[-t_{max},t_{max}]\rightarrow\mathbb{R}$.
Inspired by Bochner’s theorem~\cite{xu2019self,xu2020inductive}, the mapping function is defined as



\begin{equation}
\begin{aligned}
{\Phi}(t)=\sqrt{\frac{1}{d/2}}[cos(\omega_{1}t),sin(\omega_{1}t),\cdots,cos(\omega_{d/2}t),sin(\omega_{d/2}t)] \,,
\end{aligned}
\end{equation}
where $\omega_{1},\omega_{2},\cdots,\omega_{d/2}$ are trainable parameters.

\subsection{Gated Filter Unit}
As mentioned in the previous section, various temporal patterns are supposed to be extracted from user behaviors that are highly related to the target item. 
For such a purpose, we propose a gated filter unit that weighs each element of user behavior $S$ dynamically. 

Suppose that the target item embedding is ${e_{a}}$ and the item embedding in the user behavior ${S}$ is $\{e_{k}|1\leq{k}\leq{l}\}$. Note that item embedding dose not contain time information.
The gated weight between the target item and $k$-th behavior $s_{k}$ is expressed as 
\begin{equation}
{\rm f}_{k}=\left
\{\begin{array}{ccl}
{\rm sim}({e_{a}},{e_{k}}) && {\rm sim}({e_{a}},{e_{k}}) \geq{\delta_{thred}}\\
0 && {\rm sim}({e_{a}},{e_{k}}) < \delta_{thred} \,, \label{con:equation4}
\end{array} \right.
\end{equation}
where
\begin{equation}
{\rm sim}({e_{a}},{e_{k}})=\frac{1+\cos({e_{a}},{e_k})}{2},\ \cos({e_{a}},{e_k})=\frac{{e_{a}}{e_k}^\mathrm{T}}{|{e_{a}}|{|{e_k}|}}\,.
\end{equation}
where $\delta_{thred}$ is a hyper-parameter and can be cross validated using the test dataset for best performance. Here we use $\cos(\cdot)$ as the activation, which performs well in practice. 
Additionally, ${\rm f}_{k}$ is generated by a piece-wise function with threshold $\delta_{thred}$. 
Only when the gated weight reaches this threshold, its value remains; otherwise, it is set to $0$. Our purpose is to completely filter out the most irrelevant behaviors, which are defined as noises in this paper, and reserve the relevant behaviors with the weights ${\rm f}_{k}$.

\subsection{Time Cycle Modeling}
Together with time encoding unit and gated filter unit, we finally propose the time cycle modeling. Suppose the time embedding of $k$-th behavior $s_{k}$ is ${q_k}$ and the target time embedding of the target behavior $s_{a}$ is ${q_a}$. We aggregate the absolute time representation and relative time representation to obtain the time embeddings as
\begin{equation}
\begin{aligned}
{q_k}={\Phi}(t-t_{k})+\Lambda(t_{k}),\ {q_a}={\Phi}(0)+\Lambda(t_{a}) \,,
\end{aligned}
\end{equation}
where $t$ is the target timestamp and $\Phi(0)$ for $q(a)$ denotes that time interval is $0$.
As is shown in Figure~\ref{fig-model}, the model is divided into two modules: user interest module and time cycle module. We introduce these two modules in detail in the next subsections.

\subsubsection{User Interest Module}
The general structure of user interest module is quite similar to existing models like DIN~\cite{zhou2018deep}. The only difference is that we add time embedding to the item of user behaviors and to the target item as side information. Following the notation above, the embedding of user behavior $S$ is denoted as $\{{r_{k}}|1\leq{k}\leq{l}\}$, where ${r_{k}}={e_k} + {q_k}$, and the target embedding of the target behavior $s_{a}$ is computed by ${r_{a}}={e_{a}}+{q_{a}}$. Subsequently, the target attention is conducted between $r_{a}$ and $\{r_{k}|1\leq{k}\leq{l}\}$, which is then in the weighted sum over $r_k$ to yield the output embedding  ${r}$ of user interest module.

\begin{table}
\small
\caption{Statistics of the datasets, including the numbers of the users, items, interactions, average interactions per user (Inter. Avg. for short in the table) and train samples. 
\label{tab:stat_data}}
\begin{tabular}{l|ccccc}
\toprule
{Dataset} & {Users} & {Items} & {Inter.} & {Avg.}& {Samples}\\
\midrule
{LastFM} & {627} & {341,169} & {14,612,388} & {23,305} & {1,461,464}\\
{ML-1M} & {6,040} & {3,416} & {999,611} & {165} & {390,164}\\
{Books} & {264,522} & {954,865} & {13,616,232} & {51} & {2,577,098}\\
{IndRec} & {0.1 billion} & {1 million} & {20 billion} & {200} & {0.5 billion}\\
\bottomrule
\end{tabular}
\end{table}

\begin{table*}

\caption{
Performance comparison on four datasets. The methods of the best and the second best performance are in bold and underlined, respectively.
\label{tab:results}}
\setlength{\tabcolsep}{2mm}{
\begin{tabular}{
p{0.06\textwidth}
p{0.05\textwidth}
>{\centering\arraybackslash}p{0.05\textwidth}
>{\centering\arraybackslash}p{0.05\textwidth}
>{\centering\arraybackslash}p{0.05\textwidth}
>{\centering\arraybackslash}p{0.05\textwidth}
>{\centering\arraybackslash}p{0.06\textwidth}
>{\centering\arraybackslash}p{0.07\textwidth}
>{\centering\arraybackslash}p{0.06\textwidth}
>{\centering\arraybackslash}p{0.06\textwidth}
>{\centering\arraybackslash}p{0.06\textwidth}
>{\centering\arraybackslash}p{0.06\textwidth}}
\toprule
{Dataset} &{Metric} & {LR}& {DNN} & {DIN} & {SASRec} & {GRU4Rec} & {TimeLSTM} & {TiSASRec} & {TimelyRec} & {DiCycle} & {RelaImpr} \\
\midrule
\multirowcell{2}[0pt][l]{LastFM} &{AUC}  & {0.7353} & {0.7376} & {0.7368} & {0.7475} & {0.7389} & {0.7479}& {\underline{0.7480}}& {0.7460} & {\textbf{0.7801}}& {\textbf{12.94\%}}\\
& {GAUC}& {0.7418} & {0.7416} & {0.7432} & {0.7480} & {0.7423} & {\underline{0.7524}}& {0.7464}& {0.7416} & {\textbf{0.7961}}& {\textbf{17.31\%}}\\
\bottomrule
\multirowcell{2}[0pt][l]{ML-1M} &{AUC}  & {0.7883} & {0.8532} & {0.8561} & {0.8687} & {0.8598} & {0.8622}& {0.8648}& {\underline{0.8754}} & {\textbf{0.8955}}& {\textbf{5.35\%}}\\
&{GAUC} & {0.7962}& {0.8535} & {0.8498} & {0.8651} & {0.8601} & {0.8619}& {0.8548} & {\underline{0.8721}} & {\textbf{0.8912}} & {\textbf{5.13\%}} \\
\bottomrule
\multirowcell{2}[0pt][l]{Books} &{AUC}  & {0.7647} & {0.7802} & {0.7804} & {0.7820} & {0.7789} & {0.7833}&{\underline{0.7912}}&{0.7875} & {\textbf{0.7988}}& {\textbf{2.61\%}}\\
&{GAUC}& {0.7594} & {0.7597} & {0.7609} & {0.7546} & {0.7521}& {0.7611}& {0.7673}& {\underline{0.7691}} & {\textbf{0.7767}}& {\textbf{2.82\%}}\\
\bottomrule
\multirowcell{2}[0pt][l]{IndRec} &{AUC}  & {0.7736} & {0.7869} & {0.7930} & {0.7944} & {0.7943}& {0.7968}&{0.7979}&{\underline{0.7984}} & {\textbf{0.8046}}& {\textbf{2.08\%}}\\
&{GAUC}& {0.7198} & {0.7283} & {0.7331} & {0.7369} & {0.7352}& {0.7375}& {0.7379}& {\underline{0.7386}} & {\textbf{0.7461}}& {\textbf{3.14\%}}\\
\bottomrule
\end{tabular}}
\end{table*}

\begin{table}
\small
\caption{Ablation study. Metric AUC is adopted for evaluation. \label{tab:results_ablation}}
\setlength{\tabcolsep}{2mm}{
\begin{tabular}{ccccccccc}
\toprule
{Dataset} & {LastFM} & {ML-1M} & {Books} & {IndRec}\\
\midrule
{DiCycle}          & {\textbf{0.7801}}& {\textbf{0.8955}}&{\textbf{0.7988}}& {\textbf{0.8046}}\\
\hline
{Remove Absolute Time} & {0.7761}& {0.8906}& {0.7941}&{0.8036}\\

{Remove Relative Time} & {0.7764}& {0.8928}& {0.7932}&{0.8016}\\

{Remove Time Cycle Module} & {0.7539}& {0.8797}& {0.7862}&{0.7974}\\
\bottomrule
\end{tabular}}
\end{table}

\subsubsection{Time Cycle Module}
This module aims to learn the diverse time cycle patterns of user behaviors, which are denoised by the gated weight ${\rm f}_{k}$. Time embedding $q_{k}$ of k-th behavior $s_{k}$ turns to ${\tilde{q_k}}={\rm f}_{k}\cdot{{q_k}}$.
Some elements of $\tilde{q}=\{{\tilde{q}_{k}}|1\leq{k}\leq{l}\}$ are directly set to zero vectors by Equation ~\ref{con:equation4}, which are regarded as noises and can be harmful for time cycle modeling.

Next, the time cycle attention is applied to capture time cycle patterns based on denoised time embedding sequence ${\tilde{q}}$, where the time cycle attention is formulated as:
\begin{equation}
\begin{aligned}
c_{aj}=\frac{{q_{a}}{W_{Q}}({\tilde{q_{j}}}{W_{K}})^ \mathrm{ T }}{\sqrt d},\ \alpha_{aj}=\frac{\exp c_{aj}}{\sum_{k=1}^{l}\exp c_{ak}},\ {h}=\sum_{k=1}^{l} \alpha_{aj} ( {\tilde{q_{k}}}{W_{V}} ) \,.
\end{aligned}
\end{equation}
where $h$ is the output embedding of time cycle module and ${W_{Q}}\in \mathbb{R}^{d\times d}$, ${W_{K}}\in \mathbb{R}^{d\times d}$, ${W_{V}}\in \mathbb{R}^{d\times d}$ are parameter matrices for linear projections. Equipped with absolute time representation $\Lambda(\cdot)$ and relative time representation $\Phi(\cdot)$, the time cycle attention can find time cycle patterns that are relevant to the target item in the user behaviors. We concatenate the output embedding of user interest module ${r}$ and the output embedding of time cycle module ${h}$, then pass it into a 3-layer fully connected neural network. With activation function ${\rm Sigmoid}(\cdot)$, we obtain the probability $P(y_{j})$ that user $u$ will click on item $i$. Finally, the commonly used cross-entropy loss is chosen to train the model, which is formulated as
\begin{equation}
\begin{aligned}
\mathcal{L} = -\sum_{j=1}^{N}y_{j} {\rm log}(P(y_{j})) + (1-y_{j}){\rm log}(1-P(y_{j}))\,.
\end{aligned}
\end{equation}

%% file: experiment.tex
\subsection{Experimental Settings}
\textbf{Datasets}. We conduct experiments on three public benchmarks, including LastFM~\footnote{https://www.last.fm/}~\cite{celma2009music}, ML-1M~\footnote{https://grouplens.org/datasets/movielens/1m/}~\cite{harper2015movielens}, Amazon Books~\footnote{http://deepyeti.ucsd.edu/jianmo/amazon/index.html}~\cite{ni2019justifying}, and a real-world dataset extracted from a recommendation scenario of our company, which is named IndRec. The interactions of user and item in these datasets include click, review, and listening to music, which are uniformly termed as interaction in the following.
Statistical details of these datasets are presented in Table ~\ref{tab:stat_data}. For all the public datasets, the user-item interactions are regarded as positive samples, and negative samples are randomly sampled from the items that users have not interacted with, where the sampling rate of positive and negative samples is 1:1. We treat the last item of the behavior as test data for each user, and use the remaining for training, following ~\cite{kang2018self,sun2019bert4rec}. IndRec is extracted from two-week online logs, where samples clicked by the users are labeled as positive and samples exposed but not clicked are labeled as negative. For each sample of these datasets, the length of user behaviors is set to 200, 200, 100, 100 for LastFM, ML-1M, Books, and IndRec, respectively.


\textbf{Baselines}. Several the state-of-the-art methods serve as baselines in the experiments.
We use simple LR (Logistic Regression) and fully connected DNN (Deep Neural Network) for comparison. As for other baselines,
\textbf{DIN~\cite{zhou2018deep}} uses target attention to model user behaviors. \textbf{SASRec~\cite{kang2018self}} is a self-attention recommendation framework to model user behaviors. \textbf{GRU4Rec~\cite{hidasi2015session}} is a session-based method with structured GRU. \textbf{Time-LSTM~\cite{zhu2017next}} is a variant of LSTM equipped with time gates to model time intervals.
\textbf{TiSASRec~\cite{li2020time}} explicitly models the timestamps of interactions with self attention to explore the influence of different time intervals.
\textbf{TimelyRec~\cite{cho2021learning}} jointly learns heterogeneous temporal patterns of user preference.


\textbf{Parameters and Evaluations.}
The dimension $d$ is set to 64, the batch size is 128 and the learning rate is $1e^{-4}$ for three public datasets, and we set such numbers to 16, 1024, and $1e^{-4}$ for IndRec, respectively. We adopt two commonly used metrics to evaluate the performance, including the area under the ROC curve (AUC) and the group weighted area under the ROC curve (GAUC).
Moreover, we also report the relative improvement of DiCycle over the best baselines to show its superiority, which is measured following ~\cite{shen2021sar,yan2014coupled} by 
\begin{equation}
    RelaImpr=\left(\frac{metric(target \ model)-0.5}{metric(base\ model)-0.5}-1\right)\times{100\%},
\end{equation} where the $metric$ is AUC or GAUC.



\subsection{Results and Discussions}

\textbf{Performance}. The overall comparison results are listed in Table~\ref{tab:results}. The RelaImpr in AUC of DiCycle over the best baseline is 12.94\%, 5.35\%, 2.61\%, 2.08\% for LastFM, ML-1M, Books, and IndRec, respectively; the RelaImpr in GAUC of DiCycle over the best baseline is 17.31\%, 5.13\%, 2.82\%, 3.14\% for LastFM, ML-1M, Books, and IndRec, respectively.
The results show that DiCycle performs consistently better than all of the state-of-the-art baselines in four datasets, demonstrating its effectiveness in learning denoised temporal patterns in  user behaviors.
Moreover, an extensive ablation study is conducted with DiCycle to analyze the effect of each module, where the absolute time encoding, the relative time encoding, and the time cycle module are removed, respectively. The results in Table ~\ref{tab:results_ablation} show that all of these modules are critical to achieving final performance in the four datasets. The relative encoding has a higher effect than the absolute time encoding in Books and IndRec, whereas the opposite holds in LastFM and ML-1M. It is illustrated that the time cycle module plays a decisive role in DiCycle, without which there is a performance degradation of -9.35\%, -3.99\%, -4.22\%, and -2.36\% for LastFM, ML-1M, Books, and IndRec, respectively.

\begin{figure}[t]
\centering
\subfigure[]{
\includegraphics[width=40mm,height=29mm]{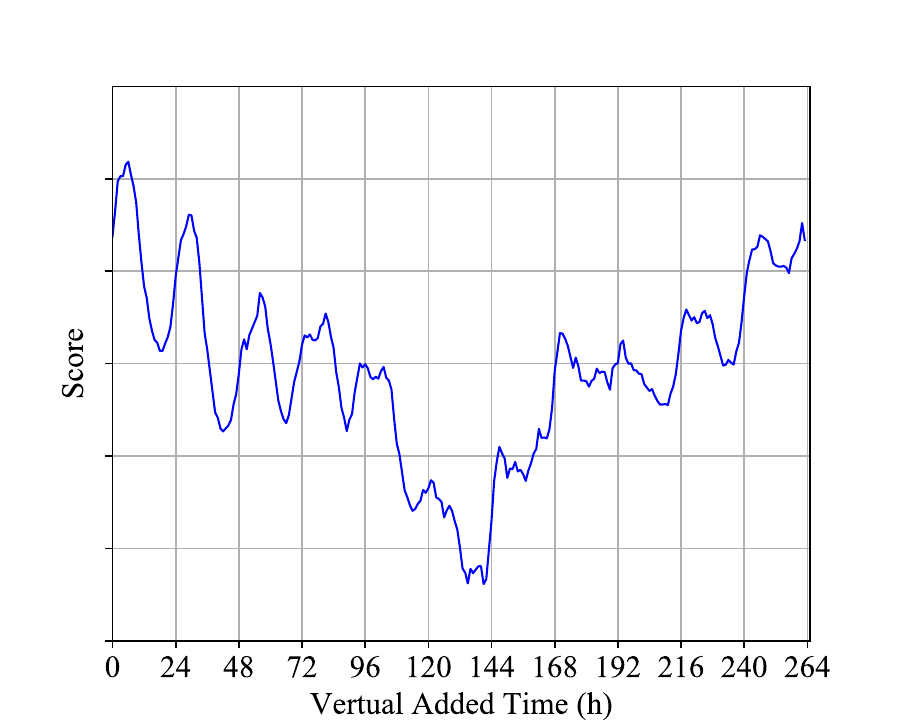}
}
\subfigure[]{
\includegraphics[width=40mm,height=29mm]{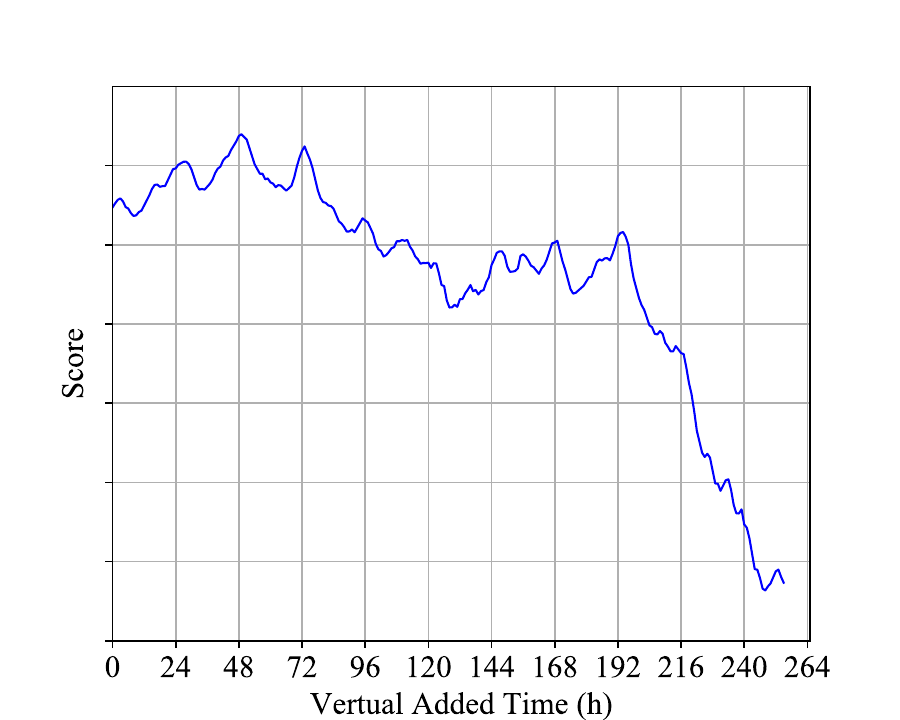}
}
\caption{(a), (b): Changing next interaction time for users who had many interactions and a few interactions with target item, respectively. \label{fig_case_study}}
\end{figure}


To provide intuitive evidence for recommendation by DiCycle, we gradually change the target timestamp given a certain user and item pair to see the variance of the prediction. As is shown in Figure~\ref{fig_case_study}, we select two interaction records from IndRec. For the first record, the user had many interactions with the target item in her past behaviors whereas very few is observed for the second one. We gradually add additional time to the target timestamp, 3600 seconds for each time. Corresponding to the first record, predicted score in terms of the CTR probability in Figure~\ref{fig_case_study}(a) show strong temporal patterns. It has a local peak every 24 hours, and the overall score range declines at first and then gets larger with the change of day. The score in Figure~\ref{fig_case_study}(b) gradually declines over time, where time cycle patterns are comparatively weaker. DiCycle filters out noises in user behaviors and can select behaviors highly related to the target time, which explains the difference between Figure~\ref{fig_case_study}(a) and Figure~\ref{fig_case_study}(b). We conclude from this case study that DiCycle has a strong capability to capture time cycle patterns for users with many interactions on items that are highly relevant to the target item.